\\

Title: Krypton-85 in the atmosphere

Authors: A. T. Korsakov, E. G. Tertyshnik

Measurement results are presented on $^{85}$Kr content in the atmosphere over the European part of Russia in 1971 – 1995 based on the analysis of samples of the commercial krypton, which is separated from air by industrial plants. Our results are by 15 % lower than $^{85}$Kr activities observed over West Europe.

According our prediction by 2030 $^{85}$Kr content in the atmosphere over Europe will amount to 1,5 – 3 Bq/m$^3$ air. Average $^{85}$Kr release to the atmosphere from regeneration of spent nuclear fuel (SNF) is estimated, some 180 TBq per 1 ton SNF.

It is advisable to recommence monitoring of $^{85}$Kr content within Russia.

Comment: 17 pages, including 3 tables, 2 figures, to be published ANRY (in Russian)

Subjects: Nuclear Experiment (nuc-ex), Geophysics (physics.geo-ph)

Journal reference:

----------------

Заголовок: Криптон-85 в атмосфере

Авторы: А. Т. Корсаков, Э. Г. Тертышник

Представлены результаты измерений содержания $^{85}$Kr в атмосфере над европейской территорией России в 1971 – 1995 гг., полученные путём анализа проб технического криптона, выделяемого из воздуха промышленными воздухоразделительными установками. Наши данные на 15 % ниже, активности $^{85}$Kr, полученной по наблюдениям над Западной Европой.

Согласно приводимому прогнозу содержание $^{85}$Kr в атмосфере над Европой к 2030 г. составит 1,5 - 3 Бк/м$^3$ воздуха. Выполнена оценка среднего поступления $^{85}$Kr в атмосферу при регенерации отработавшего ядерного топлива (ОЯТ), около 180 ТБк на тонну ОЯТ. Указывается на целесообразность возобновления мониторинга содержания $^{85}$Kr на территории России.

Комментарий: 17 страниц, включая 3 таблицы, 2 рисунка, будет опубликована в АНРИ, (на русском)

Предмет (тема): Ядерный эксперимент, Геофизика

\\

**Криптон-85 в атмосфере.**

Корсаков А.Т., Тертышник Э.Г. (ФГБУ «НПО «Тайфун»)

**Ключевые слова:** *$^{85}$Kr, атмосфера, криптон, ионизация, глобальный запас $^{85}$Kr, прогноз содержания $^{85}$Kr, регенерация ОЯТ, мониторинг активности $^{85}$Kr, метод измерения $^{85}$Kr.*

Овладение человечеством ядерной энергией сопровождается поступлением во внешнюю среду различных радионуклидов – продуктов деления урана, тория, плутония. Газообразный долгоживущий продукт деления $^{85}$Kr (период полураспада 10,76 лет) в настоящее время вносит существенный вклад в радиоактивное загрязнение атмосферы планеты. В среднем при делении 1000 тяжёлых ядер образуется 3 ядра $^{85}$Kr. Из-за химической инертности этот радионуклид является трудно улавливаемым при переработке отработавшего ядерного топлива и, поступив в атмосферу, рассеивается и сохраняется, т.к. практически не взаимодействует с подстилающей поверхностью и не вовлекается в биологические процессы. Наличие заметной радиоактивности атмосферы, обусловленной $^{85}$Kr, впервые отмечено французскими исследователями в 1954 г. [1].

В неизмеримо малых количествах $^{85}$Kr присутствует в атмосфере вследствие захвата нейтронов космического излучения атмосферным $^{84}$Kr и в меньших количествах вследствие деления природного урана. Природный криптон представляет собой смесь стабильных изотопов криптона с массовыми числами 78, 80, 82, 83, 84 и 86. Было рассчитано, что общая равновесная активность $^{85}$Kr, образованного в результате захвата нейтронов космического излучения составляет $3{,}7 \cdot 10^{11}$ Бк или 0,37 ТБк (10 Ки) [2]. Эта активность существенно меньше активности $^{85}$Kr антропогенного происхождения, которая по данным, полученным в 2010 г. немецкими исследователями [3] составляла $4{,}8 \cdot 10^{18}$ Бк или 4800 ПБк (130 МКи), причём запас $^{85}$Kr в атмосфере, образовавшегося при испытаниях ядерного оружия, составлял в 2010 г. 7,84 ПБк (0,21 МКи). В настоящее время практически весь $^{85}$Kr в земной атмосфере обусловлен выбросами промышленных предприятий, производивших оружейный плутоний, и выбросами заводов, ведущих переработку отработавшего ядерного топлива (ОЯТ).

Поскольку обмен воздушными массами между северным и южным полушариями затруднён, а основное количество радиокриптона поступает в атмосферу в северном полушарии, существует широтный ход содержания $^{85}$Kr; максимум находится в умеренных широтах северного полушария.

Для определения объёмного содержания $^{85}$Kr в атмосфере и его общего запаса измеряют объёмную активность криптона, выделенного из воздуха, поскольку состав атмосферы не изменяется ни при изменении координат земной поверхности, ни при изменении высоты. Криптон в атмосферном воздухе присутствует в очень незначительных количествах – примерно $10^{-6}$ (в каждом м$^3$ воздуха содержится 1,14 см$^3$ природного криптона). Мониторинг содержания $^{85}$Kr в воздухе, на наш взгляд, наиболее целесообразно вести путём анализа образцов технического криптона, который выделяется из воздуха с помощью промышленных установок, производящих кислород, азот, аргон и технический криптон путём низкотемпературного ожижения воздуха и последующей ректификации. Технический криптон содержит в качестве примеси 5 – 7 % ксенона. Измерение активности $^{85}$Kr в образцах технического криптона не требует использования низкофоновых установок, а результаты измерений характеризуют воздушные массы очень больших объёмов. По высоте и долготе имеет место интенсивное перемешивание воздушных потоков и распределение активности по этим координатам однородно. Для измерения зависимости объёмной активности $^{85}$Kr от широты образцы технического криптона непригодны, поскольку точки отбора проб воздуха определяются местами расположения воздухоразделительных установок. Чтобы исследовать широтный ход содержания $^{85}$Kr, используют лабораторные установки, извлекающие криптон из воздуха с помощью низкотемпературной динамической адсорбции на активированном угле или силикагеле с последующей фракционированной десорбцией и разделением фракций методом газовой хроматографии [4]. Применялись также установки с ректификационной колонной, где ожиженный воздух разделялся на лёгкую и тяжёлую фракции. Криптон, ксенон и радон концентрируются в тяжёлой фракции. На конечной стадии концентрирования использовался препаративный газовый хроматограф. Подобные установки монтируются на грузовых автомобилях или морских судах и позволяют извлекать из воздуха пробы криптона объёмом 0,5 – 1 л [5].

    Достаточно продолжительные регулярные наблюдения за содержанием в приземной атмосфере $^{85}$Kr велись в немногих странах: Франции [6], Венгрии [7] и в СССР. В последние годы опубликованы результаты измерения содержания $^{85}$Kr в Германии, Швейцарии и Японии с помощью портативных установок, извлекающих криптон из воздуха [8,9]. Над территорией США также проводились наблюдения за содержанием $^{85}$Kr в приземном слое воздуха [10] и на больших высотах [11]. В Советском Союзе регулярные измерения содержания $^{85}$Kr в приземном воздухе осуществлялись в Московском инженерно-физическом институте (МИФИ) и в Институте экспериментальной метеорологии (ИЭМ).

    Широтный ход активности $^{85}$Kr подробно исследовался учёными Франции [6] и сотрудниками МИФИ [5,12,13] с помощью установок для выделения криптона из воздуха, размещённых на судах.

    В ИЭМ измерения активности $^{85}$Kr в атмосфере были начаты в 1971 г. [14].

Отбирались пробы технического криптона, производимого кислородными установками в Новомосковске (Тульская обл., 54° с.ш.), Череповце (Вологодская обл., 59° с.ш.) и в Ереване (40° с.ш.). Пробы отбирались раз в месяц, путём перепуска криптона из стандартных 40-литровых баллонов в вакуумированные баллоны ёмкостью 0,7 л (давление 30 –50 атм.). После доставки проб в ИЭМ пробы выдерживались в течение месяца, чтобы гарантировать полный распад $^{222}$Rn (период полураспада 3,82 сут.), примеси которого могут присутствовать в техническом криптоне. После этого проба (через редуктор) подавалась в специально разработанную сцинтилляционную ячейку для определения объёмной активности $^{85}$Kr путём регистрации бета-частиц, сопровождающих его распад [15]. Для калибровки установки применяли образцовый радиоактивный газ (технический криптон, меченый $^{85}$Kr), который готовили и аттестовывали с помощью образцовых измерительных установок ВНИИФТРИ. Использование бета-счётчиков внутреннего наполнения, иногда применяющихся для измерения криптона, представляется нам нецелесообразным по ряду причин. Процентный состав газовой смеси, которая содержит измеряемый криптон, рабочий газ и гасящие добавки, необходимо строго контролировать. При введении криптона в рабочий объём счётчика возрастает вероятность появления послеимпульсов, наличие которых приводит к завышению измеряемой активности из-за большого числа метастабильных уровней у атомов криптона и ксенона. Активность некоторых образцов криптона дополнительно определяли по гамма-излучению $^{85}$Kr с помощью Ge(Li) – детектора [16]. Объём таких проб (при нормальных условиях) составлял 30 – 40 дм$^3$, продолжительность измерений достигала $5 \cdot 10^5$ с, т.к. квантовый выход гамма-квантов $^{85}$Kr (514 кэВ) крайне мал (0,434 %). Сводка среднегодовых результатов наших измерений по трём пунктам представлена в табл. 1 [17-19].

Результаты измерения отдельных проб для каждого пункта отличались от среднегодового уровня, как правило, не более чем на 15 %. Наблюдения за содержанием радиокриптона в Новомосковске велись непрерывно с 1971 по 1987 гг. Среднегодовые содержания $^{85}$Kr в Череповце и Новомосковске весьма близки, а относительное уменьшение активности на 40° с.ш. по сравнению с активностью на 54° с.ш. составляет в среднем 8 %. С 1980 г. в связи с прекращением работы воздухоразделительной установки в Ереване ежемесячный отбор и анализ проб криптона проводился в двух пунктах. В 1995 г среднегодовое содержание $^{85}$Kr для Череповца составляло 1,08 Бк на м$^3$ сухого воздуха. Заметим, что среднегодовой концентрации радиокриптона 1 Бк/м$^3$ воздуха в умеренных широтах северного полушария соответствует глобальный запас $^{85}$Kr $3,2 \cdot 10^{18}$ Бк или 3200 ПБк (86 Мки). Наши данные в среднем в 1,15 раза ниже результатов, полученных [20] для

гг.Гейделберга (Heidelberg, 42º24′ N) и Фрейбурга (Freiburg, 48º00′ N). Такое увеличение содержания $^{85}Kr$ в атмосфере над европейскими городами может быть вызвано работой предприятий по регенерации ОЯТ, расположенных в западной Европе сравнительно близко к пунктам отбора проб воздуха или различиями калибровок измерительной аппаратуры.

Многочисленные прогнозы, которые делались до аварии на Чернобыльской АЭС в 1986 г., на основе перспективных планов развития мировой ядерной энергетики, предсказывали быстрое нарастание запаса радиокриптона в атмосфере. Так, по данным Национальной комиссии по радиационной защите и измерениям США в течение нескольких десятилетий после 2000 г. содержание $^{85}Kr$ должно было возрасти до 110 Бк/м$^3$ воздуха [2]. Авторы работ [21, 22] полагали, что при таких уровнях загрязнения атмосферы радиокриптоном ионизация всей толщи атмосферы за счёт бета-частиц, испускаемых при распаде $^{85}Kr$, вызовет уменьшение электрического сопротивления атмосферы (между ионосферой и поверхностью планеты) примерно на 10 % и, вероятно приведёт к нежелательным изменениям климата. Отметим, что при объёмной активности $^{85}Kr$ в воздухе в 1 Бк/м$^3$, взаимодействие бета-частиц $^{85}Kr$ с воздухом создаёт 7,4·10$^3$ пар ионов/м$^3$ в секунду, причём ионизация воздуха происходит по всей толще атмосферы, тогда как естественный процесс ионизации за счёт распада радона и его продуктов имеет место только над поверхностью суши и в нижнем слое атмосферы, т.к. содержание радона уменьшается с высотой. Правда, другие исследователи [23, 24] считают, что при указанных уровнях

(110 Бк /м$^3$) увеличение проводимости атмосферы составит всего 1,5 % и, кроме того рекомбинация зарядов на атмосферном аэрозоле компенсирует это повышение.

Наш прогноз [18] накопления $^{85}Kr$ в атмосфере, сделанный путём экстраполяции данных наблюдений в течение ряда лет, оказался близок к реальности. В работе [18] использовались среднегодовые значения концентрации радиокриптона в Новомосковске и предположения о различных законах изменения темпов инжекции $^{85}Kr$ в атмосферу. Согласно сделанным оценкам при линейном законе увеличения инжекции со временем к 2000 г. содержание $^{85}Kr$ должно было составить 1,3 Бк/м$^3$ воздуха. Фактически среднее содержание $^{85}Kr$ в 1996 – 1998 гг. для городов ФРГ Гейделберг (Heidelberg, 49º24' N) и Фрейбург (Freiburg, 48º00' N) составило 1,4 Бк/м$^3$ воздуха [25].

На основе дополнительных данных, полученных нами в 1989 – 1995 гг. (табл. 1), и данных по содержанию в атмосфере $^{85}Kr$ над территорией Европы [25, 26] выполним экстраполяцию содержания этого загрязнителя на последующие десятилетия.

Если кроме радиоактивного распада нет других путей выведения радиокриптона из атмосферы, то изменение глобального запаса $^{85}$Kr в атмосфере можно описать уравнением

$$N(t) = N(0)\cdot\exp(-\lambda t) + \int_0^t n(t-\tau)\cdot\exp(-\lambda\tau)d\tau, \qquad (1)$$

где

$N(t)$ – глобальный запас $^{85}$Kr в атмосфере, Бк

$n(t)$ – среднегодовая инжекция, Бк/год

$t = T - T_0$, $T$ – время в годах, $T_0$ – год начала инжекции ($T_0 = 1945$)

$\lambda = \ln(2)/P$, где $P$ – период полураспада $^{85}$Kr; 10,76 лет

Полагая $N_0 = N(0)$, получим

$$N(t) = \exp(-\lambda t)\cdot(N_0 + \int_0^t n(\tau)\cdot\exp(\lambda\tau)d\tau); \qquad (2)$$

В предположении о плавном изменении глобальной годовой инжекции, для выявления доминирующей тенденции в скорости накопления $^{85}$Kr рассмотрим три гипотетических модели:

1) линейный рост $\quad n_L(t) = a_0 + a_L\cdot t$; $\quad t > 0$, $\quad a_0 > 0, a_L > 0$

2) экспоненциальный рост $\quad n_{exp}(t) = a_0 + a_1\cdot\exp(b\cdot t)$; $\quad t > 0$, $\quad b > 0$, $\quad a_0/a_1 > -1$

3) ограниченный рост $\quad n_{lim}(t) = a_0 - a_1\cdot\exp(-b\cdot t)$; $\quad t > 0$, $\quad b > 0$, $\quad a_0/a_1 > 1$

Для постоянной инжекции выражение (2) даёт

$$N_{const}(t) = N_0\cdot\exp(-\lambda\cdot t) + a_0\cdot(1 - \exp(-\lambda t))/\lambda, \qquad (3)$$

где $N_0$ – концентрация в начале измерений

$a_0$ – величина ежегодной инжекции

Тогда для выбранной формулы ограниченной инжекции можно записать

$$N_{lim}(t) = N_{const}(t) + a_1\cdot(\exp(-b\cdot t) - \exp(-\lambda\cdot t))/(b - \lambda), \qquad (4)$$

а при экспоненциальном росте инжекции

$$N_{exp}(t) = N_{const}(t) + a_1\cdot(\exp(b\cdot t) - 1)/(\lambda + b), \qquad (5)$$

Если в общем случае функция инжекции может быть представлена степенным многочленом $\quad n_m(t) = \sum_{k=0}^{m} a_k * t^k$, $\quad t > 0$,

$m$ – степень многочлена, то $\quad N_m(t) = N_0 \exp(-\lambda\cdot t) + \sum_{k=0}^{m} F_k$,

где $\quad F_0 = (1 - \exp(-\lambda t))/\lambda$;

$F_k = (t^k - k\cdot F_{k-1})/\lambda$, $\quad k = 1, 2, \ldots m$; $\quad \frac{dF_k}{dt} = k\cdot F_{k-1}$

Для случая m = 1 (линейный рост инжекции) значение среднегодовой концентрации $^{85}$Kr можно определить по формуле:

$$N_1(t) = N_{const}(t) + a_1 \cdot (t - (1 - \exp(-\lambda t))/\lambda)/\lambda,$$

и тогда зависимость концентрации в перспективе стремится к линейной:

$$N_1(t) = (a_0/\lambda - a_1/\lambda^2) + (a_1/\lambda) \cdot t$$

В общем случае легко заметить, что при установившемся монотонном характере инжекции зависимость концентрации стремится к аналогичной формуле, но с другими коэффициентами.

Таким образом, для постоянной или ограниченной инжекции (после выхода на «плато») запас радиокриптона в атмосфере стремится к величине, определяемой равновесной концентрацией:

$$N_{равн.} = a_0/\lambda \qquad (6)$$

Также как и в случае постоянной инжекции, при условии $\lambda \cdot t \gg 1$ отношение

$$n_1(t)/N_1(t)$$

стремится к величине равной $\lambda$, а величина $n_{exp}(t)/N_{exp}(t)$ при дополнительном условии $b \cdot t \gg 1$ стремится к значению $\lambda + b$.

Если предположить, начальная концентрация $^{85}$Kr в воздухе $N_0 = 0$, то наилучшее совпадение расчетной кривой с наблюдаемыми данными даёт модель с экспоненциальным ростом глобальной инжекции.

В табл. 2 представлены функции, аппроксимирующие временной ход инжекции $^{85}$Kr. Постоянные коэффициенты функций и значение $T_0$ рассчитаны методом наименьших квадратов для уравнений (3), (4), (5), причём в вариантах линейной и экспоненциальной аппроксимации использовались наши данные по среднегодовой концентрации $^{85}$Kr в г. Череповец (табл. 1), а в случае ограниченной инжекции – среднегодовые данные, полученные в западной Европе и приведённые в [3, 26, 27] для обсерватории Юнгфрауйоч (Jungfraujoch), расположенной в горах Швейцарии. Эти данные уменьшены в 1,15 раза, чтобы учесть систематическое расхождение с нашими результатами. В этой же таблице для каждого варианта изменения темпов инжекции приведены ожидаемые скорость инжекции и содержание $^{85}$Kr (умеренные широты северного полушария) в 2020 и 2030 гг. Согласно нашему прогнозу максимальная активность $^{85}$Kr в атмосфере северного полушария может составить в 2030 г. 3 Бк/м$^3$ воздуха, что примерно в 2 раза выше современного значения. На рис.1 представлены прогностические зависимости от времени содержания $^{85}$Kr в воздухе умеренных широт северного полушария, а также результаты измерений активности радиокриптона (показаны результаты измерений в России (СССР) и ФРГ). Отметим, что данные, полученные европейскими учёными [26,29] указывают на отсутствие роста активности $^{85}$Kr в атмосфере северного полушария после 2002 г., а результаты измерений, проведённых (2006 –2008 гг.) в России [28] на 10 % выше европейских и демонстрируют рост в соответствие с экспоненциальным ростом инжекции.

Согласно [2] при переработке 1 т, отработавшего ядерного топлива выделяется 130 – 1800 ТБк $^{85}$Kr. В табл. 3 приведены данные МАГАТЭ [30], из которых следует, что с 1990 по 1995 г. в мире регенерировано 10 тыс. т ОЯТ, а с 1990 по 2010 г. – 45 тыс. т. Полагая, что основная часть $^{85}$Kr, образовавшаяся в ОЯТ, поступает атмосферу в процессе регенерации, можно уточнить среднюю активность $^{85}$Kr, поступающего в атмосферу при регенерации 1 т ОЯТ. Зависимость массы регенерированного топлива от времени [30] можно аппроксимировать экспонентой (со средней относительной погрешностью менее 4 %). Дифференцируя кривую этой аппроксимации и сопоставляя полученную таким образом функцию источника с кривой инжекции, полученной по результатам измерений содержания $^{85}$Kr в Череповце (табл. 2), получаем, что средний выход $^{85}$Kr при регенерации 1 т ОЯТ составляет около 180 ТБк (рис. 2).

В последние годы содержание радиокриптона в тропосферном воздухе умеренных широт не меняется [27, 29]. Содержанию $^{85}$Kr в тропосфере умеренных широт северного полушария 1,45 Бк/м$^3$ воздуха соответствует равновесный запас этого радионуклида в атмосфере планеты $N_{равн}$ = 4,64·10$^{18}$ Бк (125 МКи), а равновесная инжекция согласно (6) $a_{равн}$ = 0,3·10$^{18}$ Бк/год (8,1 МКи/год). Согласно оценкам, приведённым в работе [31], в декабре 2001 г глобальный запас $^{85}$Kr составлял 5·10$^{18}$ Бк, причём авторы этой работы учитывали поступление воздушных масс из северного в южное полушарие, в результате которого некоторая часть $^{85}$Kr переходит из северного в южное полушарие.

Вероятно, в будущем при стабильном содержании $^{85}$Kr в атмосфере северного полушария глобальный запас этого радионуклида будет увеличиваться и содержанию $^{85}$Kr в атмосфере северного полушария (умеренные широты) 1 Бк/м$^3$ воздуха будет соответствовать глобальный запас не 3,2×10$^{18}$ Бк, как принято в наших расчётах, а увеличенное значение, которое следует уточнить, выполнив дополнительные измерения содержания $^{85}$Kr на различных широтах южного и северного полушарий.

Сокращение площади ледовых полей в Арктике вследствие глобального потепления в какой-то степени также может влиять на баланс $^{85}$Kr в атмосфере, т.к. растворимость криптона во льду меньше его растворимости в воде [32], и возрастание площади открытой воды может привести к уменьшению запаса $^{85}$Kr. С другой стороны в работе [33] отмечается, что $^{85}$Kr, сорбированный жидким абсорбентом, прочно фиксируется при переходе последнего в твёрдую фазу, поэтому благодаря этому эффекту арктические льды могут удерживать криптон, который при их таянии будет высвобождаться, увеличивая глобальный запас $^{85}$Kr в атмосфере.

Изучение широтного распределения $^{85}$Kr в южном полушарии и вблизи полюсов (от 60º до 90º) и исследование влияния площади полярных льдов на глобальный запас $^{85}$Kr требует анализа множества проб воздуха на $^{85}$Kr в течение продолжительного времени.

В перспективе эти задачи могут решаться с помощью недавно разработанного сверхчувствительного метода анализа, основанного на избирательном воздействии лазерного излучения на атомы $^{85}$Kr. Метод ATTA (Atom Trap Trace Analysis), разработанный в Аргонской Национальной лаборатории (Иллинойс, США), позволяет регистрировать отдельный атом $^{85}$Kr, который избирательно удерживается магнито-оптической ловушкой (МОЛ) и детектируется по его флуоресценции [34]. Счётная эффективность первой установки ATTA составляла $1 \times 10^{-7}$, эффективность установок, разработанных позже (ATTA-2; ATTA-3), возросла до $1 \times 10^{-4}$ [35]. Метод ATTA является относительным методом и для измерения отношения $^{85}$Kr/Kr в образце (пробе) затрачивается 2 – 3 часа, однако в процессе измерения атомы криптона, осевшие на внутренних стенках установки, искажают результаты измерения следующего образца (эффект «памяти»). Поэтому перед сменой образца приходится систему «промывать» ксеноном в течение 36 часов, в итоге производительность аппаратуры ATTA составляет всего один образец за двое суток.

В настоящее время аппаратура для ATTA анализа не получила широкого применения из-за её сложности. Три пары лазерных лучей (длина волны 811 нм) используют для удержания атомов криптона в ловушке, причём частота лазерного излучения изменяется в процессе анализа с помощью акустико-оптических модуляторов. Дополнительные лазеры применяют для замедления и фокусировки потока атомов криптона, поскольку скорость анализируемых атомов не должна превышать 20 м/с для эффективной работы МОЛ.
В районе магнито-оптической ловушки поддерживается ультравысокий вакуум ($2 \times 10^{-8}$ мм рт. ст.) посредством турбомолекулярных вакуумных насосов, работа которых дополняется применением геттеров.

Сообщается о создаваемой в Германии ATTA установке, которая позволит определять содержание $^{85}$Kr в пробе криптона, извлечённого из одного л воздуха (НТД), при времени анализа 3 часа [36]. Разработчики этой аппаратуры намерены использовать установку для обнаружения не декларированного производства оружейного плутония, т.к. отбор пробы воздуха такого объёма легко произвести в любом месте, открыв вентиль предварительно вакуумированного баллона.

Выводы:

1. Прогноз содержания $^{85}$Kr в атмосфере на 2000 г., сделанный нами в 1990 г оправдался с хорошей точностью.

2. На основе данных измерений содержания $^{85}$Kr в атмосфере, которые проводились нами на европейской территории страны с 1971 по 1995 гг., выполнен прогноз, согласно которому к 2030 г содержание $^{85}$Kr составит от 1,5 до 3 Бк/м$^3$ воздуха.

3. Приведена оценка среднего выброса $^{85}$Kr в атмосферу при регенерации 1 т ОЯТ, который составил около 180 ТБк/т.

4. Результаты определения содержания $^{85}$Kr в атмосфере над западной Европой свидетельствуют, что в последнее десятилетие прекратился рост содержания этого радионуклида, достигнув значения около 1,5 Бк/м$^3$ воздуха. При этом глобальный запас $^{85}$Kr в атмосфере составляет приблизительно 5·10$^{18}$ Бк, а инжекция $^{85}$Kr в атмосферу, компенсирующая его радиоактивный распад равна 0,3·10$^{18}$ Бк/год (8,1 МКи/год).

5. Поскольку в настоящее время на территории России не проводятся наблюдения за содержанием $^{85}$Kr в атмосфере, целесообразно возобновить мониторинг активности $^{85}$Kr в воздухе путём измерения активности проб криптоно-ксеноновой смеси, отбираемых с промышленных воздухоразделительных установок.

Т а б л и ц а 1. Среднегодовая концентрация $^{85}$Kr в приземном слое атмосферы при 0º С и 760 мм рт. ст., Бк/м$^3$ сухого воздуха (по результатам наших измерений).

| Год | Новомосковск 54º 02´с.ш. 38º 16´в.д. | Череповец 59º 08´с.ш. 37º 55´в.д. | Ереван 40º 11´с.ш. 44º 31´в.д. |
|---|---|---|---|
| 1971 | 0,50 | - | - |
| 1972 | 0,52 | - | - |
| 1973 | 0,54 | - | - |
| 1974 | 0,54 | - | 0,53 |
| 1975 | 0,60 | 0,59 | 0,54 |
| 1976 | 0,62 | 0,61 | 0,6 |
| 1977 | 0,62 | 0,62 | 0,58 |
| 1978 | 0,65 | 0,67 | 0,62 |
| 1979 | 0,62 | 0,59 | 0,53 |
| 1980 | 0,66 | 0,66 | - |
| 1981 | 0,63 | 0,67 | - |
| 1982 | 0,64 | 0,67 | - |
| 1983 | 0,72 | 0,69 | - |
| 1984 | 0,76 | 0,72 | - |
| 1985 | 0,84 | 0,79 | - |
| 1986 | 0,82 | 0,83 | - |
| 1987 | 0,91 | 0,89 | - |
| 1988 | - | - | - |
| 1989 | 0,87 | 0,96 | |
| 1990 | - | 0,93 | - |
| 1991 | - | 0,95 | - |
| 1992 | - | 0,92 | - |
| 1993 | - | 0,95 | - |
| 1994 | - | 1,02 | - |
| 1995 | - | 1,08 | - |

Т а б л и ц а 2. Варианты прогноза темпов инжекции и содержания $^{85}$Kr в атмосфере северного полушария.

| Изменение со временем среднегодовой скорости инжекции и значения расчётных коэффициентов | 2020 год | | 2030 год | |
|---|---|---|---|---|
| | годовая глобальная инжекция, **Бк / м³·год** (МКи/год) | среднегодовая концентрация, Бк / м³ возд. | годовая глобальная инжекция, **Бк / м³·год** (МКи/год) | среднегодовая концентрация, Бк / м³ возд. |
| *экспоненциальная* <br><br> **n(t)=a+b·exp(c·t)**, <br><br> a=0, b=0,0232, c=0,0296, (T0=1945 N0=0) | **6,8·10$^{17}$** (18) | **2,27** | **9,2·10$^{17}$** (25) | **3,05** |
| *линейная* <br><br> **n(t)=a$_0$+a$_1$·t**, <br><br> a$_0$=0,00557, a$_1$=0,00182, (T0=1945 N0=0) | **4,5·10$^{17}$** (12) | **1,77** | **5,1·10$^{17}$** (14) | **2,05** |
| *ограниченная* <br><br> **n(t)=a – b*exp(-c·t)**, <br><br> a=b=0,053, c=0,108, (T0=1960 N0=0,1) | **3,3·10$^{17}$** (8,9) | **1,44** | **3,4·10$^{17}$** (9,0) | **1,53** |

Т а б л и ц а 3. Количество переработанного ядерного топлива (кумулятивное количество) инжекция и содержание криптона-85 в атмосфере северного полушария.

| Годы | Масса переработанного топлива, (tHM)* | Инжекция $^{85}$Kr, Бк/год, (МКи/год) | Содержание $^{85}$Kr в Европе, Бк/м$^3$ воздуха |
|---|---|---|---|
| 1980 | - | - | 0,73 |
| 1985 | - | - | 1,04 |
| 1990 | 50000 | - | 1,06 |
| 1995 | 60000 | $0,26 \cdot 10^{18}$ (7,0) | 1,22 |
| 2000 | 75000 | $0,39 \cdot 10^{18}$ (10,5) | 1,39 |
| 2005 | 80000 | $0,13 \cdot 10^{18}$ (3,5) | 1,45 |
| 2010 | 95000 | $0,39 \cdot 10^{18}$ (10,5) | 1,45 |
| 2015 | 105000 | $0,26 \cdot 10^{18}$ (7,0) | - |
| 2020 | 124000 | $0,49 \cdot 10^{18}$ (13,2) | - |
| 2025 | 124000 | 0 | - |
| 2030 | 135000 | $0,29 \cdot 10^{18}$ (7,8) | - |

\* – тонн тяжёлых металлов (урана, плутония)

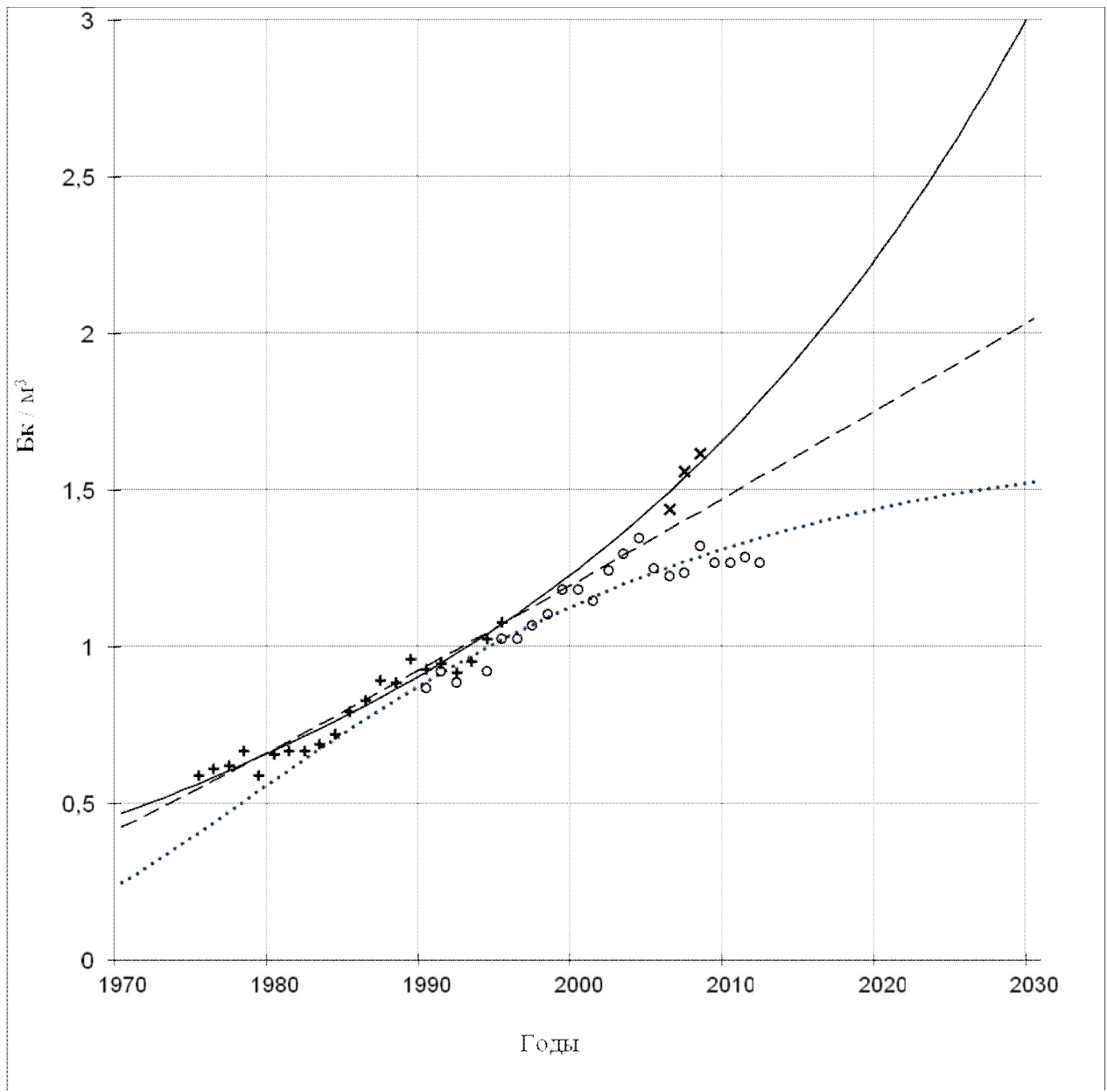

Рис. 1. Эволюция содержания $^{85}$Kr в атмосфере, (данные измерений) и прогноз для разных темпов инжекции:

⎯⎯⎯ для экспоненциальной;

− − − − для линейной;

·········· для ограниченной (табл. 2).

Данные измерений:

+ – наши результаты, г. Череповец;

× – среднегодовые результаты, приведённые в [28], г. Череповец;

○ – данные, полученные в ФРГ для обсерватории Юнгфрауйоч (Jungfraujoch), уменьшенные в 1,15 раза.

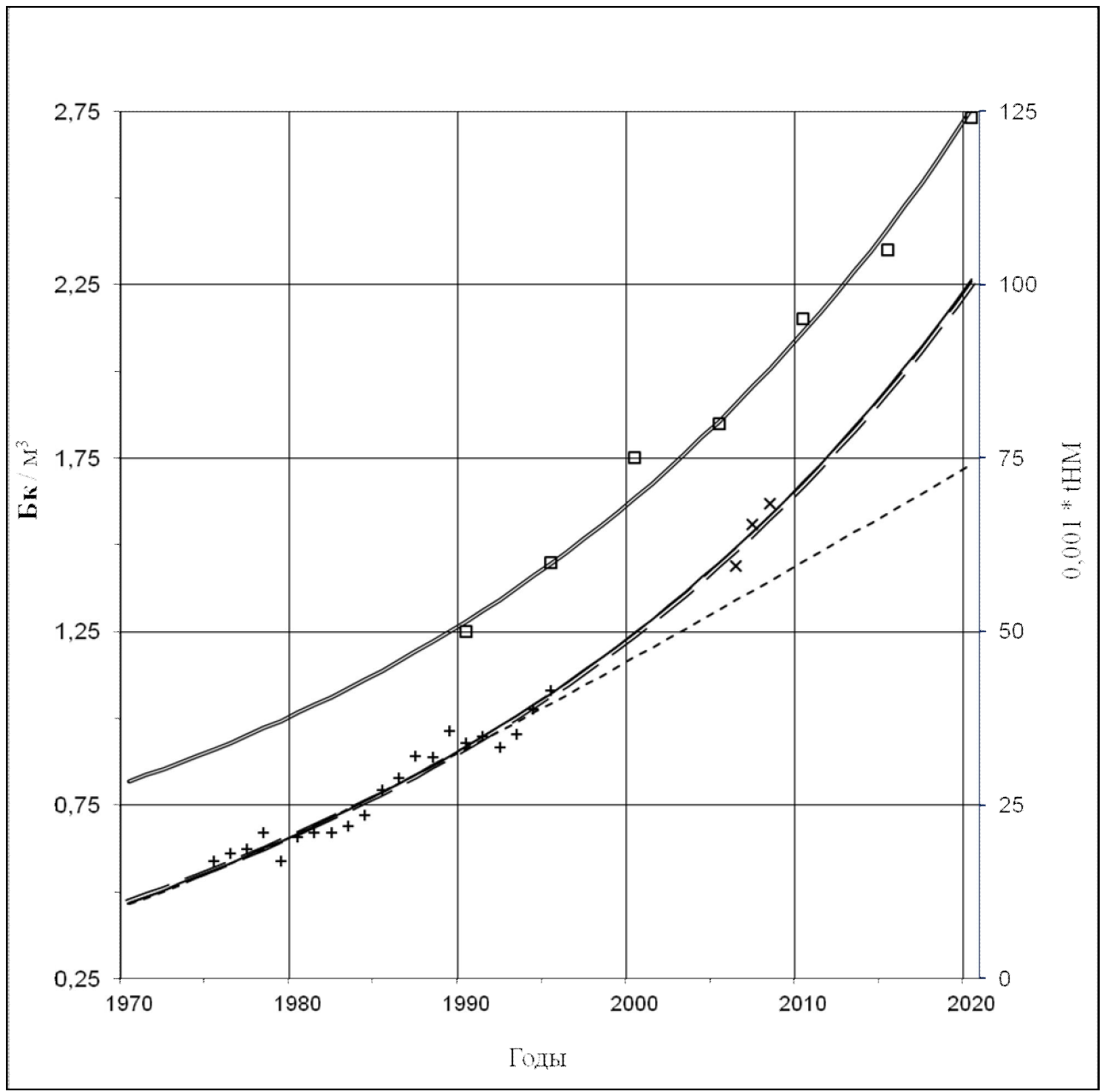

Рис. 2. К определению средней активности $^{85}$Kr, поступавшего в атмосферу при регенерации 1 т ОЯТ (177 ТБк/т) на основании данных МАГАТЭ [30] и наших измерений в г. Череповец.

□ – масса регенерированного топлива, данные МАГАТЭ;

+ – наши результаты, г. Череповец;

× – среднегодовые результаты, приведённые в [28], г. Череповец;

– – – – линейная аппроксимация инжекции

───── экспоненциальная аппроксимация инжекции

═════ экспоненциальная аппроксимация роста массы переработанного топлива

≡≡≡≡ содержание $^{85}$Kr рассчитанное, по результатам аппроксимация массы переработанного топлива